\documentclass[twocolumn,prl,aps]{revtex4}
\usepackage{graphicx}
\renewcommand{\vec}[1]{{\mathbf{#1}}}
\newcommand{\beq}{\begin{eqnarray}} 
\newcommand{\eeq}{\end{eqnarray}} 

\begin{document} 
\draft 
 
\title 
{Pseudogap in Doped Mott Insulators is the Near-neighbour Analogue of the Mott Gap}
\author{Tudor D. Stanescu}
\affiliation{Department of Physics and Astronomy, Rutgers University,
136 Frelinghuysen Rd., Piscataway, NJ., 08854-8019}
\author{ Philip Phillips}
;
\affiliation{Loomis Laboratory of Physics,
University of Illinois at Urbana-Champaign,
1100 W.Green St., Urbana, IL., 61801-3080}

\begin{abstract}
We show
that the strong coupling physics inherent
to the insulating Mott state in 2D leads to a jump in the chemical potential 
upon doping and the emergence of a pseudogap in the single particle spectrum
below a characteristic temperature.  The pseudogap arises because any singly-occupied site not immediately neighbouring a hole experiences a maximum energy barrier for transport equal to $t^2/U$, where $t$ is the nearest-neighbour hopping integral and $U$ the on-site repulsion. The resultant pseudogap cannot vanish before each lattice site, on average, has at least one hole as a near neighbour. The ubiquituity of this effect in all doped Mott insulators suggests that
the pseudogap in the cuprates has a simple origin.

\end{abstract}

\maketitle

In a Mott insulator with one orbital per site, each
unit cell contains an odd number of particles
but the Fermi energy lies in the middle of a gap. 
In contrast, band
insulators contain an even number of electrons per unit cell
and the Fermi energy lies atop a full band.  Consequently, Mott
and band insulators are {\bf  not} adiabatically connected.  Nonetheless,
most Mott insulators tend to order antiferromagnetically below some
temperature, $T_N$.  As a consequence, it is standard to view the
Mott state 
simply as an antiferromagnet in which the unit cell has doubled. On this view,
the insulating properties of a Mott insulator are equivalent
to those of a band insulator.  Band insulators possess rigid bands and hence doping only creates quasiparticles at the Fermi level.
That this picture fails fundamentally in the parent cuprates, which are all
antiferromagnetic Mott insulators, is immediately evident from optical conductivity experiments\cite{cooper,uchida} which reveal that even for $T\gg T_N$, a gap of order 2eV exists and doping leads to a massive reshuffling of spectral
weight from 2eV to the Fermi energy.  These experiments lay plain that what 
is missing in the antiferromagnetic reduction
is Mottness itself: 1) in the absence of magnetic ordering ($T>T_N$),
 a charge gap 
exists in the single particle spectrum, 2)
each electronic state in the first Brillouin zone has spectral
weight both above and below the charge gap, and 3) the sum rule that each single-particle state
carries unit weight is satisfied\cite{eskes} only when the spectral function
is integrated across the charge gap not simply up to the chemical potential as
 in a band insulator.
A consequence of (3) is that in the Mott state, the traditional notion that
 the chemical potential demarcates the boundary between
zero and unit occupancy fails fundamentally.  This failure is central to
Mottness.  

The breakdown of the traditional band insulator sum rule in the cuprates
is well described\cite{eskes} by the Hubbard model in which 
the on-site energy for double occupancy
leads to a charge gap at half-filling.
 Such local on-site physics dominates the insulating behavour at half filling. 
In the lightly doped regime, $\delta\approx 0$, effective interactions of longer range come into play
as neighbouring sites now become correlated.
If on-site correlation leads to a charge gap at half-filling, it is certainly
a possibility that nearest-neighbour correlations for $\delta\approx 0$, for example, might lead to a suppression of the density of states 
at the chemical potential as well. In fact, it is well documented that all the underdoped cuprates  
possess a pseudogap\cite{timusk} in the single-particle spectrum.  However,
the origin of this phenomenon is unknown.  As  the pseudogap does not appear to be a true $T=0$ phase and the pseudogap line
joins continuously to the Mott insulator, proposals which require
broken symmetry\cite{pg1,pg2,pg3,pg4} are difficult to
reconcile with the Mott state. 

Without any assumption as to the nature of the ground
state, we show that the the electron spectral function for the 2D Hubbard model
contains a dip at the Fermi
energy which results in a pseudogap in the single particle
 density of states (DOS).  The pseudogap remains pinned at the Fermi level
in the underdoped regime but moves above it at an intermediate doping level, 
as is seen experimentally\cite{timusk}. The pseudogap is fundamentally tied to
{\bf local} correlations on neighbouring sites much the way Mott gap arises
from on-site physics. 

The starting point for our analysis is the Hubbard model with nearest-neighbour 
hopping matrix element $t$ and on-site Coulomb repulsion $U$.  
We base our strong coupling 
analysis on a two-component composite basis $\psi$ with  
$\psi_{1\sigma}(i)=\xi_{i\sigma}$ and $\psi_{2\sigma}(i)=\eta_{i\sigma}$ and its associated Green function $S(i,j,t,t')  = 
\langle\langle \psi_{i\sigma};\psi^\dagger_{j\sigma}\rangle\rangle  =  
\theta(t-t')\langle \{\psi_{i\sigma}(t),\psi^\dagger_{j\sigma}(t')\}\rangle$,
 where $\xi_{i\sigma}=c_{i\sigma}(1-n_{i-\sigma})$ and $\eta_{i\sigma}=c_{i\sigma}n_{i-\sigma}$.  Here, $c_{i\sigma}=\eta_{i\sigma}+\xi_{i\sigma}$ annihilates an electron on site $i$ and $n_i$ is the number operator for site $i$.  
The basis $\psi$ exactly diagonalizes
the on-site interaction and hence serves as a natural starting point for 
a strong-coupling analysis.  To overcome the standard truncation
problems
inherent in the use of Hubbard operators, we adopt the following procedure.
First, project\cite{sp1} all new operators that arise from the Heisenberg equations of motion
of the Hubbard operators onto the Hubbard basis.  Second, write the 
self energy exactly in terms of the remaining operators which are now orthogonal
to the Hubbard basis.  Third, use local methods in the spirit of 
dynamical mean-field theory (DMFT)\cite{kotliar} to
calculate the resultant electron self energy.  To go beyond the single-site treatment indicative of dynamical mean-field 
theories\cite{kotliar}, we adopt the two-site expansion proposed by
Mancini and Matsumoto\cite{matsu} in which the self energy is determined self-consistently
from a two-site Hubbard cluster embedded self-consistently in an interacting
bath. As all orientations of the two-sites are considered,
the electron spectral function will be momentum dependent.
Self-consistent cluster methods which are exact 
as the limit of an infinite cluster appear to be rapidly convergent, providing
accurate results for the thermodynamics of the 1D and 2D half-filled
bands\cite{sp1} and in fact constitute
the accepted methodology\cite{kotliar} for treating strongly correlated systems.Hence, an implementation of the Hubbard operators coupled with DMFT-type technology places the limitations not on truncation in the equations of motion
but on the accuracy of the impurity solver and the size of the finite cluster. As the complete procedure is detailed
elsewhere\cite{sp1,sp2}, we mention only
that in contrast to the work of Matsumoto and Mancini\cite{matsu}, we required that 
for a fixed filling in the lattice, the chemical potential of the cluster 
equal that of the lattice. 

\begin{figure}
\begin{center}
\includegraphics[height=7cm]{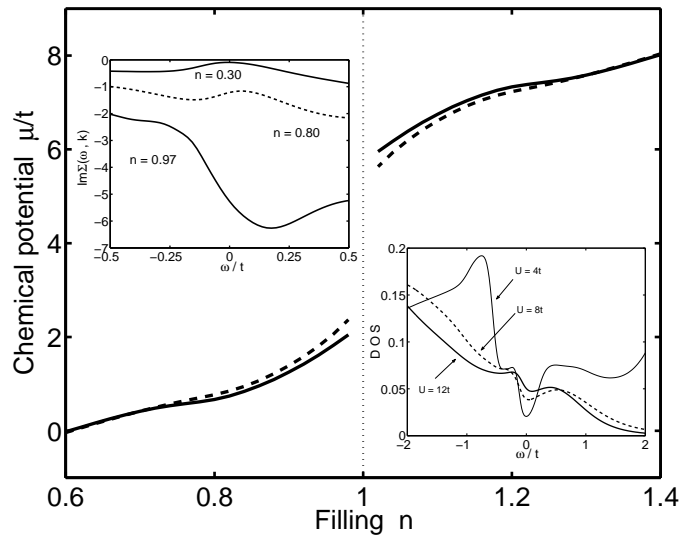}
\caption{Doping dependence of the chemical potential in the 2D Hubbard model
computed using the local cluster approach for $T=0.15t$ (dashed line)
and $T=0.07t$ (solid line).  The inset on the left shows the imaginary
part of the self energy evaluated at a Fermi momentum ($0.3,2.10$) for $n=0.97$,
($0.3,1.84$) for $n= 0.8$ and ($0.3,1.06$) for $n=0.3$, whereas
the inset on the right contains the density of states for $n=0.95$ for
$U=4t$, $U=8t$, and $U=12t$.
}
\label{fig1}
\end{center}
\end{figure}

Using this procedure\cite{sp1,sp2}, we report first the doping dependence of the chemical
potential.   Two distinct possibilities arise:  1) the chemical potential remains pinned and mid-gap states are generated or 2) the chemical potential jumps
across the Mott gap.  Our results shown in Fig. (\ref{fig1}) demonstrate that the 
chemical potential jumps upon hole or electron doping, indicating an absence 
of mid-gap states.  
The Magnitude of the jump is set by the Mott gap which
is fully developed at $T=0$.  Even for $U=4t$, the inset on the right shows that
the chemical potential resides in the LHB for $n=0.95$.
While at some
finite temperature, the chemical potential may appear to evolve
smoothly, $\Delta\mu\ne 0$ as the doping increases
and hence no mid-gap states are present.  Exact results in the 1D Hubbard model\cite{1D} as well as
Quantum Monte-Carlo simulations\cite{qmc} in 2D also reveal
a chemical potential jump upon doping and hence no mid-gap states.
However, unlike 1D and 2D, in $d=\infty$, the chemical potential remains pinned\cite{kotliar}
upon doping as 
mid-gap states emerge. 
A chemical potential jump requires a large imaginary part of the self-energy at the chemical potential. From the inset in Fig. (\ref{fig1}), we find that $\Im\Sigma$ is initially large in the underdoped regime and acquires the characteristic $\omega^2$ dependence in the overdoped regime indicative of a
Fermi liquid.  Consequently, the method we use here is capable of recovering 
Fermi liquid theory in the overdoped regime. Experimentally, whether
$\Delta\mu$ vanishes or not appears to be cuprate dependent. 
\noindent 
For example,
in La$_{2-x}$Sr$_x$CuO$_4$\cite{lsco} (LSCO), the 
chemical potential remains pinned roughly at $0.4eV$ above the top of the
LHB, while for  Nd$_{2-x}$Ce$_x$CuO$_4$ (NDCO)\cite{harima1} and
Bi$_2$Sr$_2$Ca$_{1-x}$R$_x$Cu$_2$O$_{8+y}$ (BSCO)\cite{harima2,hassan,ritveld,tjernberg}, the chemical potential jumps upon doping and scales roughly as $\delta^2$ as obtained here.  Because stripes require $\Delta\mu=0$, they have been
invoked\cite{kivelson} to explain the origin of the mid-gap states in LSCO.
The pseudogap in the underdoped cuprates has also been attributed\cite{kivelson} to stripes.
However, because $\Delta\mu\ne 0$ for both NDCO and BSCO, a requirement for 
stripe formation, if the pseudogap has a universal origin in the cuprates,
stripes are not its cause.

\begin{figure}
\begin{center}
\includegraphics[height=7.5cm]{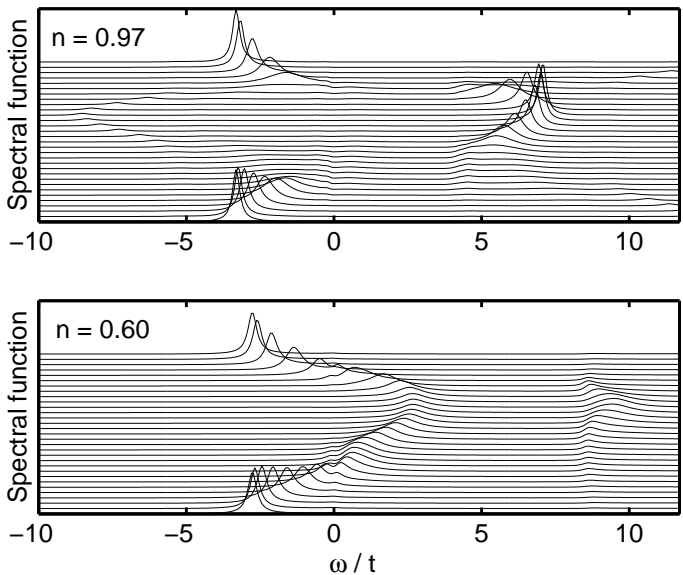}
\caption{Electron spectral function for $U=8t$, 
$T=0.07t$ and $n=0.97$  and $n=0.60$ along a path in the first 
Brillouin zone from top to bottom: $(k_x,k_y) = (0,0) \rightarrow (\pi,\pi) \rightarrow (\pi,0) \rightarrow (0,0)$.}
\label{fig2}
\end{center}
\end{figure}

To address the origin of the pseudogap, we focus on the doping 
dependence of the electron spectral function,
 $-\Im(S_{11}+2S_{12}+S_{22})/\pi)$, shown
in Fig. (\ref{fig2}).   Six features are evident:  1) the chemical
potential moves further into the LHB as the filling
decreases, 2) no coherent peaks exist near the chemical potential
for $n=.97$, 3) each state in the FBZ has spectral weight
both above and below the chemical potential as dictated by Mottness,
 4) the Mott gap remains intact but moves to higher energy as the doping
increases, 5) at $(\pi,\pi)$, the UHB carries most of
the spectral weight regardless of the filling, and 6) a dip exists
in the spectral function at the chemical potential for $n=0.97$ but
is absent for $n=0.60$.  In the underdoped regime, the characteristic
width of each $\vec k$ state is of order $t$ and even
much larger near $(\pi,0)$.  Such broad spectral features in the underdoped
regime are seen experimentally\cite{hassan} and arise in this context because
$\Im\Sigma\ne 0$ as shown in Fig. (\ref{fig1}). As a consequence, there is no sharp criterion for unit occupancy of each state in the FBZ.  Because the spectral weight at each momentum is
unity, however, and each state lives both below and above the chemical
potential, the charge carried by the piece of the state lying below the chemical
potential is less than unity.  

\begin{figure}
\begin{center}
\includegraphics[height=10cm]{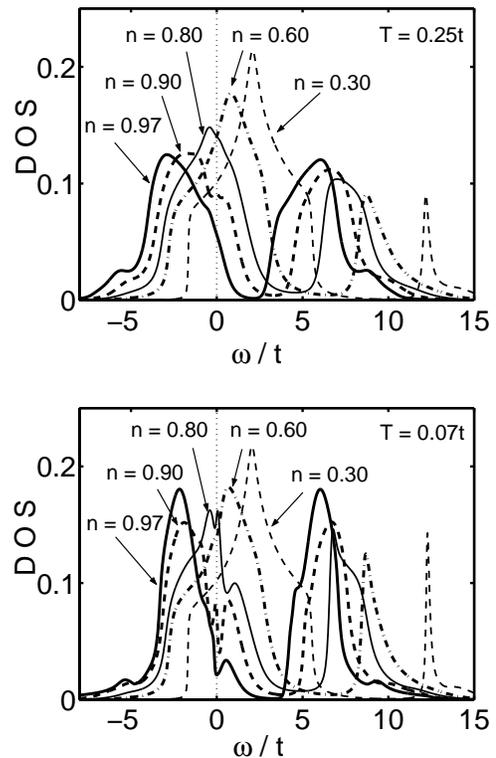}
\caption{Density of single particle states for
$T=0.25t$ and $T=0.07t$, $U=8t$ for the fillings shown.  No pseudogap
exists at high temperature.  At low $T$, a pseudogap emerges and remains pinned at the Fermi level but moves above at an intermediate doping level. In the overdoped regime, the pseudogap vanishes entirely
and a non-interacting system is recovered. }
\label{fig3}
\end{center}
\end{figure}

Is the dip in the spectral function shown in Fig. (\ref{fig2})
for $n=0.97$ real? 
Displayed in Fig. (\ref{fig3}) is the DOS for $T=0.25t$ and $T=0.07t$ for several fillings. As is evident, no local minimum of DOS exists at the chemical potential at high temperature, $T=0.25t$.  Features which emerge even at high temperature are the reshuffling of spectral
weight from above the charge gap to below as the filling is changed and also a movement of the Mott gap to higher energies.  Note that even at $n=0.30$ the Mott gap is still present, though almost all of the spectral weight now resides in the LHB
which closely resembles the non-interacting density of states.
 This is further
evidence that we correctly recover Fermi liquid theory as $n\rightarrow 0$.
At low temperature, the lower panel of Fig. (\ref{fig3}) demonstrates
that a pseudogap forms in the DOS for $\delta\approx 0$.  The vertical
line at $0$ indicates that the pseudogap occurs precisely at the chemical potential.  Similar qualitative results based on a cluster method have been obtained
by Maier, et. al.\cite{jarrell}, except their pseudogap is slightly displaced
above $E_F$.
In contrast, in the analysis of Haule, et. al.\cite{tj}, the DOS has a negative slope through $E_F$ (as dictated by the proximity to the Mott gap) but never acquires a local minimum at $E_F$ indicative of a
true pseudogap.  Because the pseudogap exists below some characteristic temperature and vanishes at higher doping, the result obtained here is non-trivial and highly reminiscent of experimental trends\cite{timusk}. 
\noindent
What is its origin?  
The inset in Fig. (\ref{fig1}) indicates that 
for a fixed filling, the pseudogap vanishes as $U$ increases and 
scales as $t^2/U$. This suggests that the pseudogap is tied to short-range correlations and hence explains why it is absent
in $d=\infty$\cite{kotliar}.  While $J_{\rm eff} \propto O(t^2/U)$
 corresponds to the energy scale for antiferromagnetic spin fluctuations, such
fluctuations cannot inhibit hole transport. In fact,
Maier, et.al.\cite{jarrell} have shown that even if antiferromagnetism
is killed, the pseudogap still persists.  Further, we have found
that $J_{\rm eff}$ is only weakly doping dependent for $0<x<.25$ and
in fact vanishes at $x\approx .8$.  Hence, the resolution
of the pseudogap problem lies elsewhere.  The energy scale, $t^2/U$ also describes
any transport process in which the intermediate state is doubly occupied. Such processes are captured by our approach as a result
of the coupling of the two-site cluster to the interacting bath.
Consider placing a single
hole in a Mott insulator. Unlike a site neighbouring the hole, a singly-occupied site two lattice sites away must temporarily doubly occupy one of its neighbours if it is to move to the hole.  The energy
 for this two-step process 
is $t^2/U$.   Sites further away experience an energy barrier with a higher power of $t^2/U$.  Hence, $t^2/U$ is the largest energy barrier for hole transport once a Mott insulator is doped.  Because some sites experience no energy barrier, the single particle density of states exhibits only a suppression, a signal that hole transport
involves virtual excitations to the UHB. This pseudogap cannot vanish before
 each site has on average one hole as its immediate neighbour, roughly $x=.25$ for a square lattice.
Hence, the pseudogap is of the form $t^2/UP(x)$, where $P(x)$ 
determines the probability that hole transport involves double occupancy and 
consequently, is a steadily decreasing function of $x$.

\begin{figure}
\begin{center}
\includegraphics[height=6cm]{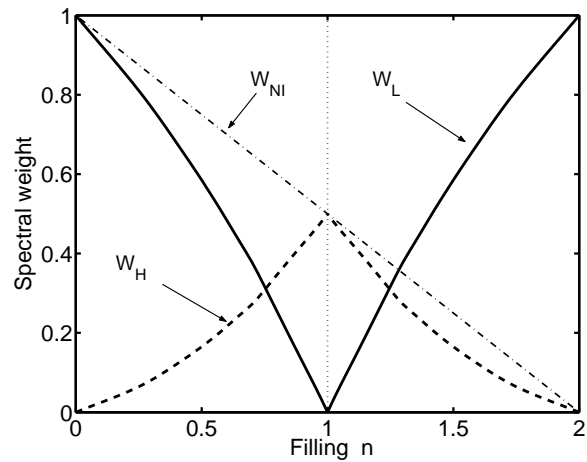}
\caption{High ($W_H$) and low ($W_L$) spectral weight as a function of filling. $W_{\rm NI}$ is the spectral
weight in the non-interacting system.}
\label{fig4} 
\end{center}
\end{figure}

Additionally, it is precisely two-step (or three-site) hopping that makes
the single-particle
low-energy spectral weight increase faster\cite{eskes,three} than $2x$.  To showthat we recover this result,
we compute
the high and low spectral weight by integrating the DOS from the energy which
minimizes the DOS to $\infty$ ($-\infty$ for electron doping) and from $\mu$ to that fixed energy, respectively.   The results shown in Fig. (\ref{fig4}) (which have been normalized per spin)
demonstrate that the initial spectral weight in the UHB which is $1/2$ at $n=1$
all moves to low energies as the filling decreases as is observed experimentally\cite{cooper,uchida}.  The same 
is true for electron doping ($n>1$).  Further, the curvature
of the low energy spectral weight (LESW) is positive indicating
that the LESW grows faster\cite{eskes,three} than $2x$. The growth in excess
of $2x$ arises entirely from virtual excitations between the LHB and UHB and 
points to an iseparability of the low and high energy scales.  Such
behavior is absent from a band insulator (see $W_{NI}$ in Fig. (\ref{fig4}). That Mottness leads to such
a drastic deviation from the non-interacting result is a direct consequence
of each state having spectral weight both above and below
the chemical potential (see Fig. (\ref{fig2})).  Our finding that the
three-site terms lead to an inseparability of low and high-energy
scales resonates with the recent work of Kirkpatrick and Belitz\cite{kb}
who have shown that three-body terms are ubiqutious in strongly
correlated electron systems and lead to breakdown of a true low-energy 
description.

Without global symmetry breaking, spin-charge separation, or pairing,
we have shown that lightly doped Mott insulators posess a pseudogap
which arises entirely from nearest-neighbour correlations.  The pseudogap
is ubituitous because any singly-occupied site not immediately neighbouring a hole experiences an energy gap for transport equal to $t^2/U$.  This generic
phenomenon offers a simple
resolution of the pseudogap problem in the cuprates. 

\acknowledgements This work was funded by the Petroleum Research Fund of the
ACS.


\begin{thebibliography}{99}
\bibitem{cooper}S. L. Cooper, et. al., Phys. Rev. B {\bf 41}, 11605 (1990).
\bibitem{uchida}S. Uchida, et. al. Phys. Rev. B {\bf 43}, 7942 (1991).
\bibitem{eskes}M. B. J. Meinders, H. Eskes, and G. A. Sawatzky,
 Phys. Rev. B {\bf 48}, 3916-3926 (1993).
\bibitem{timusk}T. Timusk and B. Statt, Rep. Prog. Phys. {\bf 62}, 61 (1999).
\bibitem{pg1}V. J. Emery and S. A. Kivelson, Nature {\bf 374},
434 (1995); M. Randeria, Varenna Lectures, cond-mat/9710223 (1997).
\bibitem{pg2}C. M. Varma, Phys. Rev. Lett. {\bf 83}, 3538 (1999).
\bibitem{pg3}S. C. Zhang, Science {\bf 275}, 1089 (1997).
\bibitem{pg4}S. Chakravarty, et. al.,
cond-mat/0005443; D. A. Ivanov, 
P. A. Lee, X.-G. Wen, Phys. Rev. Lett. {\bf 84}, 3958 (2000).
\bibitem{sp1}T. D. Stanescu and P. Phillips, Phys. Rev. B, {\bf 64}
 235117/1-8 (2001).
\bibitem{kotliar}D. S. Fisher, G. Kotliar, and G. Moeller, Phys. Rev. B {\bf
52} 17112 (1995).
\bibitem{matsu}H. Matsumoto and F. Mancini, Phys. Rev. B {\bf 55}, 2095 (1997).
\bibitem{sp2}T. D. Stanescu and P. Phillips, cond-mat/0301254.
\bibitem{1D}F. Woynarovich, J. Phys. C {\bf 15},
 85 (1982); {\bf 15}, 97 (1982).
\bibitem{qmc}M. Jarrell and T. Pruschke, Phys. Rev. B {\bf 49}, 1458 (1993).
\bibitem{lsco}G. Rietveld, M. Glastra, and D. van der Marel,
Physica C {\bf 241} 257 (1995); A. Ino,  Phys. Rev. Lett. {\bf 79}, 2101 (1997).
\bibitem{harima1}N. Harima, et. al., cond-mat/0103519.
\bibitem{harima2}N. Harima, A. Fujimori, T. Sugaya, and I. Terasaki,
cond-mat/0203154.
\bibitem{hassan}M. Z. Hassan, et. al., Science {\bf 288}, 1811 (2000).
\bibitem{ritveld}G. Ritveld, S. J. Collocot, and D. van der Marel, 
Physica C, {\bf 241} 273 (1995)
\bibitem{tjernberg}M. A. van Veenendaal, et. al., Phys. Rev. B {\bf 47}, 446 (1993).
\bibitem{kivelson}E. W. Carlson, V. J. Emery, S. A. Kivelson, D. Orgad,
cond-mat/0206217.
\bibitem{jarrell}Th. A. Maier, M. Jarrell, A. Macridin, and F.-C. Zhang,
cond-mat/0208419.
\bibitem{tj}K. Haule, A. Rosch, and P. W\"olfle, cond-mat/0205347.
\bibitem{pines}A. P. Kampf and J. R. Schrieffer, Phys. Rev. B {\bf 42}, 7967 (1990); D. Pines, Physica C {\bf 282-287}, 273 (1997).
\bibitem{three}H. Eskes, A. M. Ole\'s, M. B. J. Meinders, and 
W. Stephan, Phys. Rev. B {\bf 50}, 17980 (1994).
\bibitem{kb}T. R. Kirkpatrick and D. Belitz, cond-mat/0303151.
\end{thebibliography}
\end{document}